# MOLECULAR DYNAMICS STUDY OF POLARIZATION EFFECTS ON AgI

*Vicente Bitrián and Joaquim Trullàs**

Departament de Física i Enginyeria Nuclear, Universitat Politècnica de Catalunya, Campus Nord UPC, 08034 Barcelona, Spain.

* Corresponding author. E-mail: quim.trullas@upc.edu

Three different models of AgI are studied by molecular dynamics simulations. The first one is the rigid ion model (RIM) with the effective pair potential of the Vashishta and Rahman form and the parameterization proposed by Shimojo and Kobayashi. The other two are polarizable ion models in which the induced polarization effects have been added to the RIM effective pair potential. In one of them (PIM1) only the anions are assumed to be polarizable by the local electric field. In the other one (PIM2s) the silver polarization is also included, and a short-range overlap induced polarization opposes the electrically induced dipole moments. This short-range polarization is proved to be necessary to avoid overpolarization when both species are assumed to be polarizable. The three models reproduce the superionic character of $\alpha$-AgI at 573 K and the liquid behavior of molten AgI at 923 K. The averaged spatial distribution of the cations in the $\alpha$-phase obtained for PIM1 appears to be in better agreement with experimental data analysis. The PIM1 also reproduces the structure factor prepeak at about 1 $\text{Å}^{-1}$ observed from neutron diffraction data of molten AgI. The three models retain in the liquid phase the superionic character of α-AgI as the mobility of the cations is significantly larger than that for the anions. The ionic conductivity for the polarizable ion models is in better agreement with experimental data for *a*-AgI and molten AgI.

## 1. Introduction

Silver iodide (AgI) is one of the most extensively studied superionic conductors, namely systems characterized by having exceptionally high (liquid-like) values of ionic conductivity *s* in the solid state. This high conductivity is reached because of the diffusive motion of one of the constituent species through the sublattice formed by the particles of other species.[1] At room temperature and pressure AgI consists of a mixture of the *g* and *b* phases, which have the cubic zinc-blende and the hexagonal wurtzite structure, respectively.[2] The *b*-phase becomes the more stable phase above 410 K, and a first-order structural phase transition to the superionic *a*-phase takes place at 420 K accompanied by an increase in *s* of around three orders of magnitude. In the *a*-phase the iodides form a body centered cubic (bcc) lattice, while the silver ions diffuse through it. It has been shown experimentally, by means of neutron[3,4,5] and x-ray[3,6] diffraction, and extended x-ray absorption fine structure studies,[7] that the $Ag^+$ ions occupy predominantly the tetrahedral sites located on the faces of the bcc unit cells, with the diffusion between these sites occurring primarily along [110] directions via trigonal sites. The high *a*-AgI ionic conductivity is related to this high degree of disorder in the $Ag^+$ positions (there are 12 available tetrahedral sites and only 2 cations in each unit cell). The conductivity at the phase transition temperature (420 K) is *s* = 1.3 $(\Omega cm)^{-1}$, a value comparable to that of molten salts, it increases with temperature to reach *s* = 2.6 $(\Omega cm)^{-1}$ at the melting point (825 K), and decreases by approximately 10% on melting.[8]

Computer simulation has been a powerful tool in the study of AgI properties, both in solid and molten phases. In the first molecular dynamics (MD) study Vashishta and Rahman[9] introduced a polynomial form for the effective pair interionic potential parameterized using crystal data. They obtained good agreement with experiment for the calculated properties of *a*-AgI. Later, Parrinello, Rahman and Vashishta[10] carried out MD simulations in which the volume and shape of the simulating cell were allowed to change with time. With the same functional form but a different parameterization from that in ref 9, their effective pair potential (which we will refer to as PRV potential) was capable of describing the *b*↔*a* transition. The PRV potential was also used by Tallon[11] to simulate a phase diagram of silver iodide similar to that observed for actual AgI; by Chiarotti et al.[12] to study the motion of $Ag^+$ ions in the *a*-phase; and by Madden et al.,[13] who suggested the ordering tendency of $Ag^+$ ions observed in simulation as the mechanism responsible for the *a*→*b* transition.

The PRV potential has also been used in simulations of the molten phase. Howe et al.,[14] as a result of a diffuse neutron scattering experiment, noted the similarity between the structure of *a*-AgI and that of molten CuCl, which melts from a superionic phase. This result prompted Stafford and Silbert[15] to suggest that, since the PRV potential was suitable for a description of the superionic *a*-phase, it should also be a good potential choice to describe molten AgI. They carried out theoretical calculations of the radial distribution functions and structure factors within the hypernetted chain approximation (HNC). Shortly after this work, Takahashi et al.[16] published the first neutron diffraction data for the coherent static structure factor of molten AgI and found a qualitative agreement between theoretical and experimental results. Moreover, MD simulations using the PRV potential suggested that molten AgI, near melting, retains somehow the superionic character of the *a*-phase.[17,18] Later, Shimojo and Kobayashi[19] modified slightly the parameterization of the PRV potential in order to reproduce the *a*-phase features at the appropriate experimental density (larger than in ref 10).

The ionic induced polarization was not taken into account in the above simulation works. However, Madden and Wilson[20] showed the important role of polarization in some ionic systems, like alkaline earth halides, and stated that it can account for phenomena often attributed to a certain degree of covalency. They proposed two different mechanisms for the induction of ionic polarization: a purely electrostatic induction governed by a fixed polarizability, and a mechanical short-range polarization that depends on the nearby ions. This latter contribution, that opposes the former, allows for the fact that the dipole moments created by the field are damped at short interionic separations where overlap effects become significant.

The influence of the induced polarization on the properties of other silver halides such as AgCl and AgBr has already been investigated. Wilson et al.[21] simulated solid and liquid AgCl by using a model in which dipole and quadrupole polarization effects were added to a pair potential of the Born-Mayer form. Their results for the optic phonon dispersion curves and the melting temperature were in much better agreement with experiment than those obtained with a simple pair potential, and they reproduced qualitatively the distinctive three-peak feature in the main broad peak of the experimental structure factor. Later, Trullàs et al.[22] showed that the three-peak structure of molten AgCl can also be reproduced if only the anion induced dipole polarization contributions are added to a pair potential of the form proposed by Vashishta and Rahman in ref 9. Furthermore they found that, unlike ref 21, the results for the conductivity were in good agreement with experimental values. Recently, we carried out a MD study on molten AgBr by using a rigid ion model and a polarizable ion model in which only the anionic induced polarization is considered.[23] The latter reproduces the broad main peak of the structure factor obtained by neutron diffraction experiments, although the smoothed three-peak feature of this broad peak is slightly overestimated. We also studied a model in which both species were polarizable but a non-physical overpolarization ('polarization catastrophe') took place. In order to solve this problem, according to refs 20 and 21, we added a short-range damping contribution to the dipole moment that opposes the dipole induced by the electric field created by the ionic charges. However, the polarization catastrophe was not avoided. We saw that simulations could be stable longer than $10^5$ time steps, but for some improbable but possible critical configurations two unlike ions became over-polarized.

The purpose of this work is twofold. Firstly, we propose a new model that avoids the polarization catastrophe when both species are polarizable. In this model the short-range damping contribution opposes the dipole induced by the electric field created by both the ionic charges and the induced dipoles. The second aim is to complete the paper series about the polarization effects on the structure and ionic transport properties of molten silver halides. We have already studied molten AgCl and AgBr in previous papers[22,23], and we have published recently a preliminary MD study on the structure of molten AgI and the origin of the prepeak observed in the experimental structure factor.[24] In this recent communication we used a polarizable ion model in which the anions are only polarizable by the electric field. Now, in the present work, we have also carried out MD simulations of a new model in which both cations and anions are polarizable. Furthermore, we have also checked that the ionic models reproduce some well-known characteristics of the solid *a*-phase.

The layout of the paper is as follows. We describe the models in section 2. In section 3 we describe the computational details. In section 4 we present and discuss the results of our simulations. Finally, we summarize our results in the concluding remarks of section 5.

**2. Interaction models**

**2.1. The rigid ion model.** The potential energy of the rigid ion model (hereafter referred as RIM) is

$$U^{\text{RIM}} = \sum_{i=1}^{N-1} \sum_{j>i}^{N} f_{ij}^{\text{VR}}(r_{ij}), \quad (1)$$

where $N$ is the number of ions, and $r_{ij}$ is the distance between two ions. The functional form of the effective pair potential is that originally proposed by Vashishta and Rahman,[9] which has been widely used to describe silver and copper halides (for a review see ref 1),

$$f_{ab}^{\text{VR}}(r) = f_{ab}^{0}(r) - \frac{P_{ab}}{r^4}, \quad (2)$$

with

$$f_{ab}^{0}(r) = \frac{z_a z_b e^2}{r} + \frac{H_{ab}}{r^{h_{ab}}} - \frac{C_{ab}}{r^6}, \quad (3)$$

where we use the $a$ and $b$ subscripts to denote species Ag$^+$ or I$^-$ rather than particles. The first term on the right hand side of eq. (3) is the Coulomb interaction between ionic charges, with $|z_a| < 1$ the effective charge in units of the fundamental charge $e$; the second models the repulsion between the ions arising from the overlap of the outer shells of electrons; and the third is the van der Waals contribution. From recent ab initio MD simulations of molten AgI, Shimojo et al.[25] have found that the valence electrons of Ag ions are not completely transferred to I ions, what supports the use of effective charges $|z_a| < 1$. The last term in eq. (2) is the effective monopole-induced dipole attractive interaction, with $P_{ab} = \frac{1}{2}(a_a z_b^2 + a_b z_a^2)e^2$ where $a_a$ are the electronic polarizabilities. This is the way in which polarization terms are approximated in the RIM as two-body terms, ignoring the many-body nature of induced polarization. We have used the parameterization proposed by Shimojo and Kobayashi (see Table 1).[19] At this point it is worth noting that in the original PRV potential $H_{ab} = A(s_a + s_b)^{h_{ab}}$ where $s_{Ag} = 0.53$ Å and $s_I = 2.20$ Å are related to the ionic radii.[10] Although Shimojo and Kobayashi modified slightly the values of $H_{ab}$, their parameterization keeps the large size difference between cations and anions.

TABLE 1: Potential parameters of the RIM, eq. (2), for AgI, with $|z_a| = 0.5815$.

|  | $Ag^+$–$Ag^+$ | $Ag^+$–$I^-$ | $I^-$–$I^-$ |
|---|---|---|---|
| $h_{ab}$ | 11 | 9 | 7 |
| $H_{ab}$ / eVÅ$^{h_{ab}}$ | 0.16 | 1310 | 5328 |
| $P_{ab}$ / eVÅ$^4$ | 0 | 14.9 | 29.8 |
| $C_{ab}$ / eVÅ$^6$ | 0 | 0 | 84.5 |

In the polarizable ion models $P_{ab} = 0$ for all interactions.

**2.2. The polarizable ion models.** The polarizable ion models are constructed by adding the many-body induced polarization interactions to the pair potential $f_{ab}^0(r)$ of eq. (3) with the same parameter values given in Table 1. We assume that two types of point dipoles are induced in an ion placed at position $r_i$. The first one is the dipole induced by the local electric field $E_i$ due to all other ions, whose moment is given by the polarizability $a_i$ in the linear approximation $a_i E_i$. The second type is the deformation dipole induced by short-range overlap effects due to the neighboring ions.[26] It can be physically interpreted as a correction for the fact that the electronic distribution is not well represented by point charges and dipoles at small distances.[27] For the resulting dipole moment $\mathbf{p}_i$ we use the following constitutive relation

$$\mathbf{p}_i = a_i \mathbf{E}_i + \mathbf{p}_i^s, \tag{4}$$

where the superscript $s$ stands for "short-range damping". The local electric field at $\mathbf{r}_i$ is

$$\mathbf{E}_i = \mathbf{E}_i^q + \mathbf{E}_i^p = \sum_{j \neq i}^N \mathbf{E}_{ij}^q + \sum_{j \neq i}^N \mathbf{E}_{ij}^p, \tag{5}$$

where

$$\mathbf{E}_{ij}^q = \frac{z_j e}{r_{ij}^3} \mathbf{r}_{ij} \quad \text{and} \quad \mathbf{E}_{ij}^p = 3 \frac{\mathbf{p}_j \cdot \mathbf{r}_{ij}}{r_{ij}^5} \mathbf{r}_{ij} - \frac{\mathbf{p}_j}{r_{ij}^3} \tag{6}$$

are the electric field at $\mathbf{r}_i$ created by the charge $q_j = z_j e$ and by the induced dipole on the ion $j$ (within the point dipole approximation), respectively. If $\mathbf{p}_i^s$ is approximated to be additive, a general constitutive relation for $\mathbf{p}_i^s$ can be written as

$$\mathbf{p}_i^s = -a_i \sum_{j \neq i}^N [f^q(r_{ij}) \mathbf{E}_{ij}^q + f^p(r_{ij}) \mathbf{E}_{ij}^p], \tag{7}$$

where $f^q(r)$ and $f^p(r)$ are suitable damping functions only effective over length scales corresponding to the nearest-neighbor separation and limiting value 1 as $r \to 0$. By choosing this form we ensure that the short-range terms cancel the dipole moment induced by the electric field if two ions approach unphysical separations. Notice that Eq (7) could have been written in a more general form with $f_{ij}^q(r)$ and $f_{ij}^p(r)$ instead of $f^q(r)$ and $f^p(r)$, but this is not the case in this work. The relation suggested by Madden and coworkers,[21,26] and that we used in ref 23, is recovered from eq. (7) with $f^p(r) = 0$. This relation works well for a variety of divalent and trivalent metal halide models in which only the anions are assumed to be polarizable.[28] However we found that the polarization catastrophe takes place for silver halide models in which both species are polarizable. This is the reason why in this work we also include the damping of the dipole moment induced by the $\mathbf{E}_{ij}^p$ terms, in a similar way as other authors (see, for example, refs 29 and 30). In this work we have chosen the simplest extension of the model used in ref 23 by assuming

$$f^q(r) = f^p(r) = f(r) = \exp(-r/\boldsymbol{r}) \sum_{k=0}^{4} \frac{(r/\boldsymbol{r})^k}{k!}, \tag{8}$$

where $f(r)$ is the Tang and Toennies dispersion damping function.[31,26] The parameter $\boldsymbol{r}$ determines the length scale over which the damping acts, but we will characterize $f(r)$ by $\boldsymbol{l} = 4.671\boldsymbol{r}$, the distance at which $f(\boldsymbol{l}) = 0.5$. The short-range polarization effects are eliminated in the limit $\boldsymbol{l} = 0$. Substitution of eqs. (5)-(8) in eq. (4) leads to a system of linear equations that determines uniquely the $N$ dipole moments $\{\mathbf{p}_i\}$ for given ionic positions $\{\mathbf{r}_i\}$.

The general expression of the potential energy for the above polarizable ion models can be written as

$$U^{\mathrm{PIM}} = \sum_{i=1}^{N-1} \sum_{j>i}^{N} f_{ij}^0(r_{ij}) - \sum_{i=1}^{N} \mathbf{p}_i \cdot \mathbf{E}_i^q - \frac{1}{2} \sum_{i=1}^{N} \mathbf{p}_i \cdot \mathbf{E}_i^p + \frac{1}{2} \sum_{i=1}^{N} \frac{\mathbf{p}_i^2}{\boldsymbol{a}_i} + U^s, \tag{9}$$

where

$$U^s = \sum_{i=1}^{N} \mathbf{p}_i \cdot \sum_{j \neq i}^{N} f^q(r_{ij}) \mathbf{E}_{ij}^q + \frac{1}{2} \sum_{i=1}^{N} \mathbf{p}_i \cdot \sum_{j \neq i}^{N} f^p(r_{ij}) \mathbf{E}_{ij}^p. \tag{10}$$

The first term on the right hand side of eq. (9) corresponds to the potential energy of a rigid ion model whose ions interact via the pair potential $f_{ij}^0(r)$. The second, third and fourth terms give the usual potential energy of point dipoles linearly induced by the electric field.[27,32,33]. The last term is due to the inclusion of the short-range damping polarization. It is obtained by imposing that, given an ionic configuration, the minimum of the potential energy occurs when the values of the dipole moments satisfy eq. (4), i.e., the gradient of $U^{\mathrm{PIM}}$ with respect to each dipole moment vanishes. The fact that $\tilde{\mathbf{N}}_{\mathbf{p}_i} U^{\mathrm{PIM}} = 0$ (for $i = 1, ..., N$) implies that all contributions to the forces coming from terms involving derivatives with respect to the dipole moments are zero. Then, only the explicit spatial derivatives of $U^{\mathrm{PIM}}$ have to be considered, and the force acting on the ion $i$ can be written as in ref 22 plus an extra term due to the negative gradient of $U^s$, whose expression is

$$\mathbf{F}_i^s = \mathbf{F}_i^{sq} + \mathbf{F}_i^{sp}, \tag{11}$$

$$\mathbf{F}_i^{sq} = \sum_{j \neq i}^N \left( z_j e \left[ f^q(r_{ij}) \mathbf{E}_{ji}^p - \frac{\dot{f}^q(r_{ij})}{r_{ij}^4} (\mathbf{p}_i \cdot \mathbf{r}_{ij}) \mathbf{r}_{ij} \right] - z_i e \left[ f^q(r_{ij}) \mathbf{E}_{ij}^p - \frac{\dot{f}^q(r_{ij})}{r_{ij}^4} (\mathbf{p}_j \cdot \mathbf{r}_{ij}) \mathbf{r}_{ij} \right] \right), \tag{12}$$

$$\mathbf{F}_i^{sp} = -\sum_{j \neq i}^N \left( f^p(r_{ij}) \tilde{\mathbf{N}}_{\mathbf{r}_i} [\mathbf{p}_i \cdot \mathbf{E}_{ij}^p] + \frac{\dot{f}^p(r_{ij})}{r_{ij}} (\mathbf{p}_i \cdot \mathbf{E}_{ij}^p) \mathbf{r}_{ij} \right), \tag{13}$$

where $\dot{f}^q(r) = df^q(r)/dr$, $\dot{f}^p(r) = df^p(r)/dr$. Eq. (12) is equivalent to the extra term given in ref 23 and eq. (13) corresponds to the new short-range damping contributions introduced in this work.

In this work we present the results of two polarizable ion models. In both models, the anions are assumed to be polarizable with $a_I = 6.12$ Å$^3$ (ref 19). In the first one, which we denote PIM1, the cations are not polarizable ($a_{Ag} = 0$) and the dipole moments of the anions are induced only by the local electric field ($l = 0$). Then $f(r) = 0$ and the terms related to the damping effects, i.e., $\mathbf{p}_i^s$ in eq. (4), $U^s$ in eq. (9) and the force contribution given by eq. (11), vanish. In the second model, which we denote PIM2s, the cation polarizability is $a_{Ag} = 1.67$ Å$^3$ (ref 34), and $l = 1.84$ Å. This $l$ value has been chosen according to the relationship $l = (s_{Ag} + s_I)/c$ proposed in ref 28, with the value of $c$ derived from the $l$ used for AgCl in ref 21.

**2.3. The potential energy of two isolated ions.** Substitution of eqs. (5)-(8) in eq. (4) leads to a system of linear equations that can be solved analytically for the two-particle case, and it is found that the potential energy of two isolated polarizable ions can be written as

$$f_{12}^{PIM}(r) = f_{12}^0(r) - [1 - f^q(r)]^2 \frac{\frac{1}{2}\left([a_1 z_2^2 + a_2 z_1^2] r^3 - z_1 z_2 r_c^6 [1 - f^p(r)]\right) e^2}{r(r^3 + r_c^3[1 - f^p(r)])(r^3 - r_c^3[1 - f^p(r)])}, \tag{14}$$

where $r_c = (4a_1 a_2)^{1/6}$. It is worth noting that eq. (14) with $a_1 = a_{Ag} = 0$, $a_2 = a_I \neq 0$ and $f^q(r) = f^p(r) = 0$, as for PIM1, becomes equal to the effective pair potential $f_{AgI}^{VR}(r)$ of eq. (2) proposed for the RIM. Of special interest is the case of two unlike polarizable ions. In Figure 1 we plot $f_{12}^{PIM}(r)$ for a cation and an anion with $a_1 = a_{Ag} = 1.67$ Å$^3$, $a_2 = a_I = 6.12$ Å$^3$ and different damping functions. If $f^q(r) = f^p(r) = 0$, $f_{12}^{PIM}(r)$ presents a divergence at $r_c = (4a_1a_2)^{1/6}$ because the induced dipole moments become infinity at this *polarization catastrophe distance*. MD simulations of the whole system under these conditions polarize catastrophically whatever the initial configuration is. If $f^q(r) = f(r)$ with $l = 1.84$ Å and $f^p(r) = 0$, as we did in ref 23 for AgBr, $f_{12}^{PIM}(r)$ also diverges at the same critical distance $r_c$, but now there is a maximum at a longer $r$. Therefore, the total force between two unlike ions that approach each other is repulsive at distances slightly larger than $r_c$. MD simulations of the bulk under these conditions show that for some improbable but possible configurations an ion has a velocity large enough to overcome this energy maximum, whose form in each configuration depends significantly on the many-body effects. Although these simulations can be stable for a long time (more than $25 \times 10^3$ time steps), they also polarize catastrophically. However, if $f^q(r) = f^p(r) = f(r)$ with $l = 1.84$ Å, as for PIM2s, $f_{12}^{PIM}(r)$ does not exhibit any divergence. Hence, the conclusion drawn from this two-particle analysis, and confirmed by MD simulations of the whole system, is that the damping of the dipole moment contribution due the field created by the other dipoles must be considered in order to avoid the overpolarization. Given $r_c = (4a_{Ag}a_I)^{1/6} = 1.86$ Å, the value of $l$ must be larger than 1.6 Å, because at lower values the factor $(r^3 - r_c^3[1 - f^p(r)])$ in the denominator of eq. (14) vanishes at a certain $r$ and the divergence is not avoided.

In the case of two like polarizable ions, i.e., $z_1 = z_2$ and $a_1 = a_2$, the factor $(r^3 - r_c^3[1 - f^p(r)])$ in the denominator of eq. (14) also appears in the numerator and cancels itself, and the potential energy for two ions of species $a$ can be written as

$$f_{aa}^{PIM}(r) = f_{aa}^0(r) - [1 - f^q(r)]^2 \frac{a_a z_a^2 e^2}{r(r^3 + 2a_a[1 - f^p(r)])}, \qquad (15)$$

without a polarization catastrophe distance even when $f^q(r) = f^p(r) = 0$ as for PIM1. However, MD simulations for a similar model of molten NaI with only the anions polarizable have shown that two anions can approach at distances at which they polarize catastrophically,[35] even though this is not predicted for two isolated anions. Notice that eq. (15) with $a_1 = a_2 = a_I \neq 0$ and $f^q(r) = f^p(r) = 0$, as for PIM1, does not become equal to the effective pair potential $f_{II}^{VR}(r)$ of eq. (2).

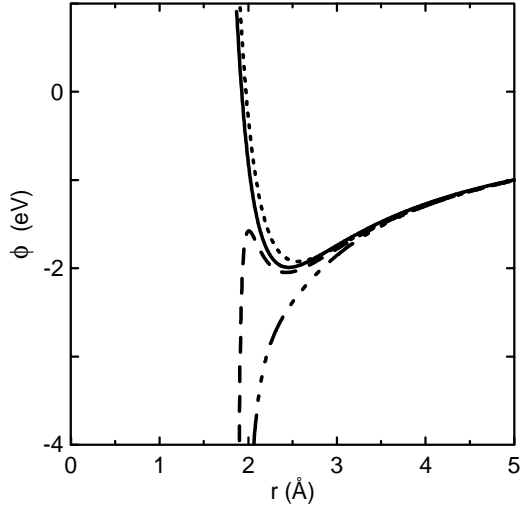

**Figure 1.** Potential energy $f_{12}^{PIM}(r)$ given by eq. (14) for two isolated unlike polarizable ions with $a_1 = a_{Ag} = 1.67$ Å$^3$, $a_2 = a_I = 6.12$ Å$^3$, the pair potential parameters of Table 1, and the following damping functions: $f^q(r) = f^P(r) = 0$ (dash-3-dots line); $f^P(r) = 0$ and $f^q(r) = f(r)$ with $l = 1.84$ Å (dashed line); and $f^q(r) = f^P(r) = f(r)$ with $l = 1.84$ Å as for PIM2s (dotted line). The solid line corresponds to the potential energy of two unlike ions with $a_1 = a_{Ag} = 0$, $a_2 = a_I = 6.12$ Å$^3$ and $f^q(r) = f^P(r) = 0$, as for PIM1, which is equal to $f_{AgI}^{VR}(r)$ given in eq. (2).

### 3. Computational details

By using the potential models described above, we have simulated solid AgI at 573 K, the temperature at which Cava et al.[4] carried out neutron diffraction measurements, and molten AgI at 923 K, the same at which recent neutron diffraction data are available.[36] The corresponding ionic number densities $r_N$ are 0.0306 ions/Å$^3$ (ref 5) and 0.0281 ions/Å$^3$ (ref 37), respectively.

MD simulations have been carried out with $N = 500$ ions placed in a cubic box of side $L$ with periodic boundary conditions, and using the Beeman's algorithm with a time step of $\Delta t = 5 \times 10^{-15}$ s. The electric fields $\mathbf{E}_i^q$ and $\mathbf{E}_i^p$, and the corresponding long range interactions between charges and induced dipole moments, have been evaluated by the Ewald method. The Ewald sums used in our simulations are given in ref 22. The Ewald parameters are the same as those used in ref 23. Notice that $f^q(r_{ij})\mathbf{E}_{ij}^q$ and $f^p(r_{ij})\mathbf{E}_{ij}^p$ are short-range contributions and, thus, can be evaluated without the Ewald method. We have also carried out MD simulations of the RIM and PIM1 with 1000 ions, and we have verified that all the results are practically the same as those obtained with 500 ions. However, the time required in the PIM2s case with 1000 ions is 25 times longer than that for RIM (see below), beyond our present computing time capability.

The system of equations that results from substituting eqs. (5)-(8) into eq. (4), whose solution are the dipole moments for given ionic positions, may be solved by using the matrix inversion method. It is based on the matrix equation[27,29]

$$\mathbf{A} \cdot \mathbf{p} = \mathbf{E}', \qquad (16)$$

where $\mathbf{p} = (\mathbf{p}_1,...,\mathbf{p}_N)$ and $\mathbf{E}' = (\mathbf{E}'_1,...,\mathbf{E}'_N)$ are 3$N$-component vectors with $\mathbf{E}'_i = \Sigma_{i \neq j} \mathbf{E}_{ij}^q [1-f^q(r_{ij})]$. $\mathbf{A}$ is a 3$N \times$3$N$ matrix that can be split into 3$\times$3 submatrices $\mathbf{A}_{ij}$ ($i,j = 1,...,N$) with elements

$$(\mathbf{A}_{ij})_{mn} = d_{ij} d_{mn} \frac{1}{a_i} - (1 - d_{ij})[1 - f^p(r_{ij})](\mathbf{T}_{ij})_{mn}, \qquad (17)$$

where $d_{ij}$ is the Kronecker delta and $\mathbf{T}_{ij}$ is the dipole field tensor,

$$\mathbf{T}_{ij} = 3\mathbf{r}_{ij}\mathbf{r}_{ij}/r_{ij}^5 - \mathbf{I}/r_{ij}^3, \qquad (18)$$

with $\mathbf{I}$ the identity matrix. Inverting the matrix $\mathbf{A}$ we find, from $\mathbf{p} = \mathbf{A}^{-1} \cdot \mathbf{E}'$, the dipoles for a given configuration. However, this exact method is very expensive computationally and it is more convenient to calculate the dipole moments by using the prediction-correction iterative method proposed by Vesely.[38,22] In this iterative method, an initial guess for the fields $\{\mathbf{E}_i^p\}$ and for the dipoles $\{\mathbf{p}_i\}$ is made by using the dipole moments from the previous time step, and the dipole moments resulting from these fields and dipoles are evaluated using eq. (4), which can be iterated to self-consistency.

In the PIM1 case we have used the iterative method in all the time steps with a convergence limit of $\left|\Delta\mathbf{E}_i^p\right|^2 / \left|\mathbf{E}_i^p\right|^2 < 10^{-4}$ for each ion. The convergence is reached after about 15 iterations and the computer time spent is about 1.4 times longer than in the RIM case. In the PIM2s case we have found that the

iterative process does not converge for a few percent of the time steps, as it leads to unstable oscillations in the predicted dipoles. Then, we have used the iterative method when possible and the matrix inversion when the former does not work. With $N = 500$, the iterative procedure converges after about 100 iterations in the 99.9% of the time steps, but the matrix inversion is needed for the other 0.1%, and the computer time spent is about 12 times longer than in the RIM case. With $N = 1000$, the convergence is reached after about 300 iterations in 98% of the time steps, the matrix inversion is needed for the other 2%, and the computer time spent was about 25 times longer than in the RIM case, beyond our present computing time capability.

All simulations have been started by placing the anions on a bcc lattice and two cations per unit cell at two of the tetrahedral sites. Anyway, we have checked that all the equilibrium properties obtained for the solid do not change if cations are placed in any other position of the anionic lattice. The solid properties have been averaged over $500 \times 10^3$ equilibrium configurations. On the other hand, in order to get configurations in which the anions exhibit liquid behavior, we started the simulations of the molten phase at an ionic number density smaller and at a temperature higher than the density and temperature of interest. Once equilibrium was reached, we compressed and cooled the system in several steps, making sure that at each step the liquid behavior was preserved, until reaching the density and temperature of interest. Then, the liquid properties have been averaged over $10^6$ equilibrium configurations. This large number minimizes the statistical fluctuations in the structure factors and reduces the uncertainties in the values of the conductivity.

The basic structural properties calculated in the simulations are the three partial radial distribution functions, $g_{ab}(r)$, and the corresponding Ashcroft-Langreth structure factors, $S_{ab}(k)$.[39,40] We have used the hybrid method described in ref 23 for the calculation of $S_{ab}(k)$. From $S_{ab}(k)$ and the coherent scattering lengths $b_{Ag} = 5.922$ fm and $b_I = 5.280$ fm,[41] the coherent static structure factor, normalized in such a way it goes to 1 as $k \to \infty$, is given by $S(k) = [b_{Ag}^2 S_{AgAg} + b_I^2 S_{II} + 2 b_{Ag} b_I S_{AgI}]/(b_{Ag}^2 + b_I^2)$.[42]

Regarding the ionic transport properties, the self-diffusion coefficients, $D_a$, and ionic conductivity, $\sigma$, have been calculated via Kubo integral of velocity correlation functions and Einstein relation for the slope of mean square displacements, in the same way as in ref 35. The estimated uncertainties for the $D_a$ values are less than 1%, and that for $\sigma$ is less than 5%. The accordance between the values calculated using the Kubo and the Einstein relations is within the error intervals. As in previous papers,[22,23] the conductivity has been calculated using the formal charge $|z_a| = 1$. Whereas in their interactions we assume the ions only "see" effective charges, in their transport we assume the ions carry with them the full complement of electrons.

## 4. Results

**4.1. α-phase structure.** In order to check that the three models reproduce the superionic behavior of the α-phase, we have carried out MD simulations at 573 K starting from a configuration in which the iodides occupy bcc sites. In the three cases the iodides remain around lattice sites, with a mean square displacement around 0.6 Å$^2$. On the other hand, the silver ions show a liquid-like behavior independently on their initial positions and diffuse through the essentially rigid framework formed by the anions. A detailed analysis of the ionic transport properties is presented in subsection 4.3.

In the three model simulations of α-AgI the silver ions tend to be located on the faces of the cubic bcc unit cells the most of the time. There is an iodide site at each one of the vertices of this face. In Figure 2 the normalized silver ion density distribution for PIM1 in one of these faces is shown. This density distribution has been calculated by averaging the positions of the silver ions within a plate of thickness $a/16$, $a = 5.075$ Å being the lattice parameter.[5] Since the density maps for the other two models are very similar we do not present them. There are higher-density occupancy regions around the four tetrahedral sites and, to a lesser extent, around trigonal and octahedral sites (see the caption of Figure 2). This result is in agreement with the experimental work of Cava et al.[4] and our results are consistent with previous studies (see ref 1) which concluded that the diffusive motion of silver ions consists of jumps between adjacent tetrahedral sites via trigonal sites, with a small occupation of octahedral sites.

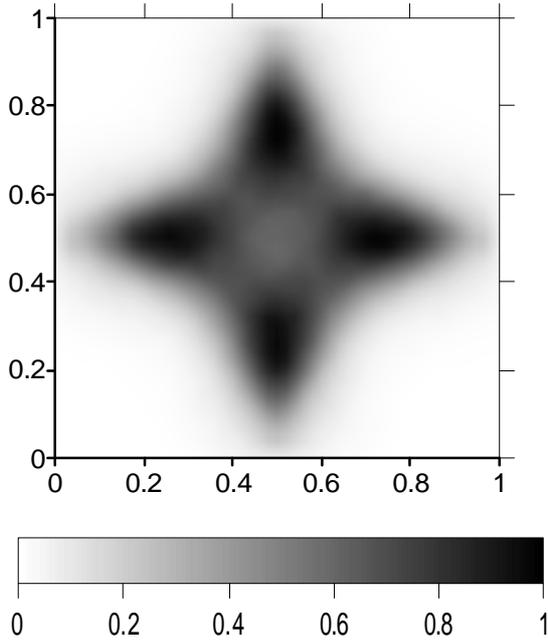

**Figure 2.** Normalized silver ion density distribution in a (100) face of the conventional bcc cubic unit cell of $\alpha$-AgI from MD simulations at 573 K using PIM1. The lengths represented on the axis are normalized by the lattice parameter $a$. There is an iodide site at each one of the vertices of the face. The sites at (0.5$a$, 0.25$a$), (0.25$a$, 0.5$a$), (0.5$a$, 0.75$a$), and (0.75$a$, 0.5$a$) are tetrahedral sites; those at the middle of the segment between two adjacent tetrahedral sites are trigonal sites; and those at the center of the face and at the middle of its four sides are octahedral sites.

We have also calculated the partial radial distribution functions, $g_{ab}(r)$. $g_{AgI}$ and $g_{II}$ do not present significant differences from one model to another. The positions of the $g_{II}$ peaks correspond to the distances between the sites of a bcc lattice with lattice parameter $a$. The I-I coordination number (namely, the average number of iodides that are at distances smaller than that of the first $g_{II}$ minimum from a given iodide) is around 14, as in all bcc structures at high temperatures (8 nearest and 6 next-nearest neighbors). The first peak of $g_{AgI}$ is at 2.7 Å, slightly smaller than the distance between an iodide site and a tetrahedral site (2.8 Å).

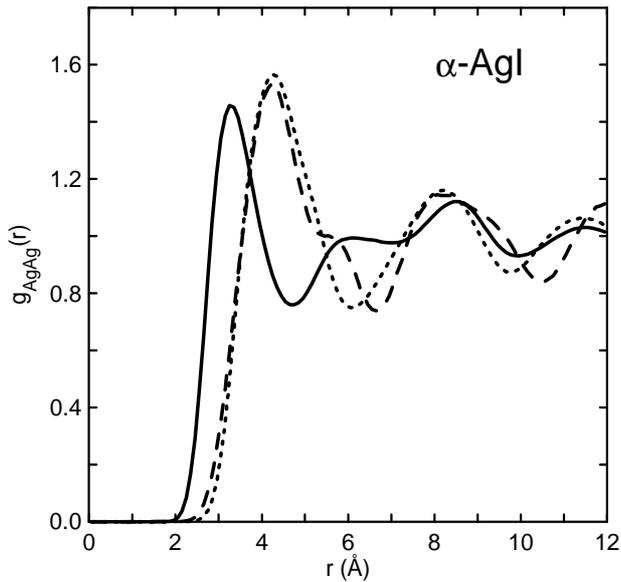

**Figure 3.** Cation-cation radial distribution function, $g_{AgAg}(r)$, from MD simulations of *a*-AgI at 573 K using the RIM (dotted line), PIM1 (solid line), and PIM2s (dashed line).

On the other hand, the $g_{AgAg}$ shown in Figure 3 present interesting differences. The most striking one is that the first peak of the $g_{AgAg}$ for PIM1 is shifted backward at 3.2 Å. This shift, which has also been observed in the liquid phase (see below) and for the polarizable model of molten AgBr,[23] can be attributed to the screening of the repulsion between cations due to polarization. The negative ends of the anion dipoles attract the positive ions and, therefore, the separation between the cations can be smaller than it would be the case if the anions were not polarized. As it was already suggested by Gray-Weale and Madden,[43] the PIM1 second peak (or shoulder) at 6.10 Å can be due to those pairs of neighboring cations attracted by different anion dipoles. Then, these two peaks can be interpreted as the result of the splitting of the RIM first peak at 4.3 Å because of the inclusion of polarization. In the PIM2s case, although the position of the first peak is very close to that for RIM because of the short-range damping effects, there is a shoulder at about the position of the PIM1 second peak. The shift of the $g_{AgAg}$ first peak found for PIM1 is similar to that found by Chahid and McGreevy for the superionic *a*-phase of CuI at high temperature and high pressure.[44] The $g_{CuCu}(r)$ that they obtained from Reverse Monte Carlo (RMC) modeling of experimental scattering data exhibits a first peak at a lower $r$ than that found in MD simulations using a rigid ion pair potential.

It is also interesting to relate $g_{AgAg}(r)$ to the occupation of the tetrahedral sites of the bcc cubic unit cells. In the three cases there is a zero probability of finding silver ions closer together than 1.8 Å, which excludes the simultaneous occupation of two nearest tetrahedral sites. The position of the first peak of $g_{AgAg}$ for RIM and PIM2s (4.3 Å) is approximately the distance between two tetrahedral sites linked by the vector $(1/2,1/2,1/2)a$. From all possible distributions of two cations per unit bcc cell at the tetrahedral sites, this is the maximum nearest-neighbor separation. However, the first peak for PIM1 is at 3.2 Å, which means that two nearest cations tend to locate around tetrahedral sites linked by $(1/2,1/4,1/4)a$ (or equivalent vectors of the same modulus). The RMC analysis performed by Nield et al.[5] and the first-principles MD calculations carried out by Wood and Marzari[45] also exhibited the first peak of $g_{AgAg}$ at around 3 Å. Nevertheless, they obtained the second peak at approximately the same position as the first peak of our RIM and PIM2s $g_{AgAg}$. Their results, then, show the closer approach between neighboring silver ions found for PIM1, but also include the correlations at ~4.3 Å found for RIM and PIM2s.

**4.2. Liquid structure.** Figure 4 shows the partial radial distribution functions obtained from liquid-state simulations. Some characteristics are common to the three models: (a) each ion is surrounded by a shell of unlike nearest-neighbor ions; (b) $g_{AgI}$ oscillates in antiphase with $g_{II}$; and (c) the silver ions, because of their smaller size, show a less marked structure and a deeper penetration into the first coordination shell of a like ion. However, while the $g_{AgAg}$ for RIM oscillates in phase with $g_{II}$, the first peak of $g_{AgAg}$ for PIM1 and PIM2s lies between the peaks of $g_{AgI}$ and $g_{II}$, and the oscillations beyond it are strongly smoothed. For PIM1 and PIM2s, since the induced dipoles on the iodides screen the repulsion between $Ag^+$ ions, the neighboring cations can approach at shorter distances, as it has been found for the *a*-phase.

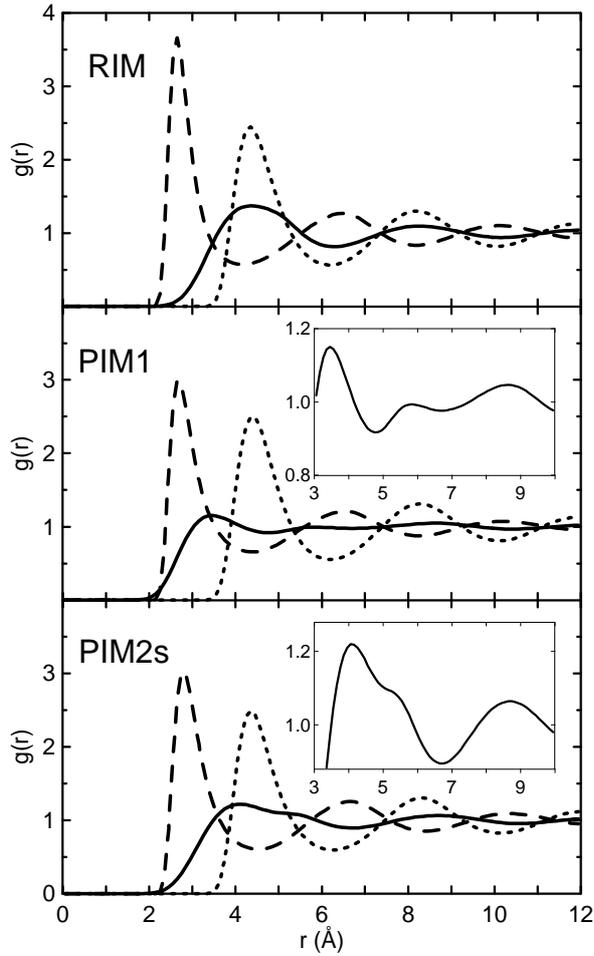

**Figure 4.** Partial radial distribution functions, $g_{AgAg}(r)$ (solid line), $g_{II}(r)$ (dotted line), and $g_{AgI}(r)$ (dashed line), from MD simulations of molten AgI at 923 K using the RIM (top), PIM1 (middle), and PIM2s (bottom). The insets show $g_{AgAg}(r)$ in more detail.

The PIM1 $g_{AgAg}$ for the liquid follows the same trends as that obtained for the *a*-phase. It has the first peak at a shorter distance (3.45 Å) than that for RIM (4.35 Å), and presents a small second peak at 5.85 Å (see the inset in Figure 4). Polarization effects are less intense for PIM2s, because of the short-range damping, and the first peak of the $g_{AgAg}$ is in between those for RIM and PIM1, at 4.10 Å, with a shoulder at a position close to that of the second PIM1 peak (see the inset). It is worth noting that the first peak of the $g_{AgAg}$ for liquid PIM2s is at a shorter distance than that for the *a*-phase.

The $g_{II}$ for the three models are almost identical. The iodides, because of their large size, are so close-packed that their structure does not depend significantly on the details of the pair potential or on the induced polarization effects. Nevertheless, comparison between the three $g_{II}$ shows little differences. The first peak of $g_{II}$ for PIM1 is slightly higher (2.49) and at a larger position (4.40 Å) than that for RIM (2.45 and 4.35 Å), with that for PIM2s in between. These little differences are probably related to the fact that the pair potential $f_{II}^0(r)$ used for PIM1 and PIM2s does not include the attractive contribution $-P_{II}/r^4$ added in $f_{II}^{VR}(r)$ for RIM. Therefore the repulsion between iodides is slightly weaker for RIM and they are slightly less packed. It is as if the effective iodide size is slightly smaller for RIM.

Although the $g_{AgI}$ for the three models are very similar, they present the following differences. The first peak of the $g_{AgI}$ for RIM is at 2.65 Å and it is higher than those for PIM1 at 2.70 Å and PIM2s at 2.85 Å. These differences strengthen the idea that effective iodide size is slightly smaller for RIM. The inclusion of the effective monopole-induced dipole attraction term $-P_{AgI}/r^4$ in the pair potential $f_{AgI}^{VR}(r)$ for RIM shifts its minimum at a shorter distance (2.46 Å) than that of $f_{AgI}^0(r)$ at 2.56 Å. The attractive force between the iodide dipoles and silver cations is also included in the polarizable models via the many body interactions added to $f_{AgI}^0(r)$. However, despite the fact that the potential energy $f_{AgI}^{PIM}(r)$ of two isolated unlike ions for the PIM1 case is equal to $f_{AgI}^{VR}(r)$, it appears that the reduction of the effective iodide size due to the changes on the pair potentials is slightly more pronounced than that due to the induced many body interactions.

Some of the structural features found for PIM1 and PIM2s have also been observed in experimental studies of molten copper halides. The absence of a marked cation structure in molten CuCl was determined by Eisenberg et al.[46] from experiments of neutron diffraction with isotopic substitution. In their RMC study on molten CuBr, which also melts from a superionic phase, Nield et al.[47] found the peak of $g_{CuCu}$ at a lower $r$ than that obtained from MD simulations using a rigid ion pair potential. They attributed this shift to polarization effects. Shirakawa et al.[48] deduced the partial radial distribution functions of molten CuX (X=Cl, Br and I) from neutron diffraction data and they found that, for the three cases, the $g_{CuCu}$ first peaks lie between those of $g_{CuX}$ and $g_{XX}$.

The partial structure factors $S_{ab}(k)$ are shown in Figure 5. The charge ordering is reflected in the reciprocal space by the fact that the main peaks of $S_{AgAg}$ and $S_{II}$, and the valley of $S_{AgI}$, are at the same wave-number $k_M \approx 1.7$ Å$^{-1}$. As expected from the $g_{ab}(r)$, there are not appreciable differences in the three $S_{II}$, and only slight ones in $S_{AgI}$. There is a little displacement of the first minimum and the first maximum of $S_{AgI}$, with the shortest $k$ for the PIM2s and the longest for the RIM, that is, the inverse order of the $g_{AgI}$ peaks. The most salient differences are those exhibited by $S_{AgAg}$. The main peak at $k_M$ of the $S_{AgAg}$ for PIM1 is much lower than that for PIM2s, which in turn is lower than that for RIM. The $S_{AgAg}$ for PIM1 presents a second peak at 2.25 Å$^{-1}$ related to the second peak of the corresponding $g_{AgAg}$ at 5.85 Å. The distance between this second peak and the third one at 8.65 Å of the $g_{AgAg}$ for PIM1 (see Figure 4) is 2.80 Å, whose associated wavelength is ($2p$/2.80Å) = 2.24 Å$^{-1}$. The $S_{AgAg}$ for PIM2s presents intermediate features between those for RIM and PIM1. It also exhibits a second maximum after the principal peak, but lower than that for PIM1. Nevertheless, the most striking polarization effect is the prepeak exhibited at $k_1 \approx 1$ Å$^{-1}$ by the $S_{AgAg}$ for PIM1. In ref 24 we discussed its structural origin and suggested that it is the signature of a periodicity of low cation density zones (or voids) due to anion polarization effects. The shift to lower $r$ of the $g_{AgAg}$ first peak for PIM1 with respect to RIM implies the existence of regions of high cation density. As a consequence of this inhomogeneous Ag$^+$ distribution, the voids open up, giving rise to a new intermediate-range characteristic length scale, signaled by the prepeak. While this prepeak was not found in previous HNC calculations or MD simulations of molten AgI using pairwise potentials,[15,17,19] it has been obtained in a recent ab initio MD study by Shimojo et al.[25] In Figure 5(a) we compare the $S_{AgAg}$ for PIM1 with that by Shimojo et al., as well as that obtained by Kawakita et al.[49] after RMC analysis of neutron diffraction data. Our MD $S_{AgAg}$ is similar to that by Shimojo et al. However, comparison between the corresponding $g_{AgAg}$ shows some differences similar to those between our *a*-phase $g_{AgAg}$ and those by Nield et al.[5] and by Wood and Marzari.[45]

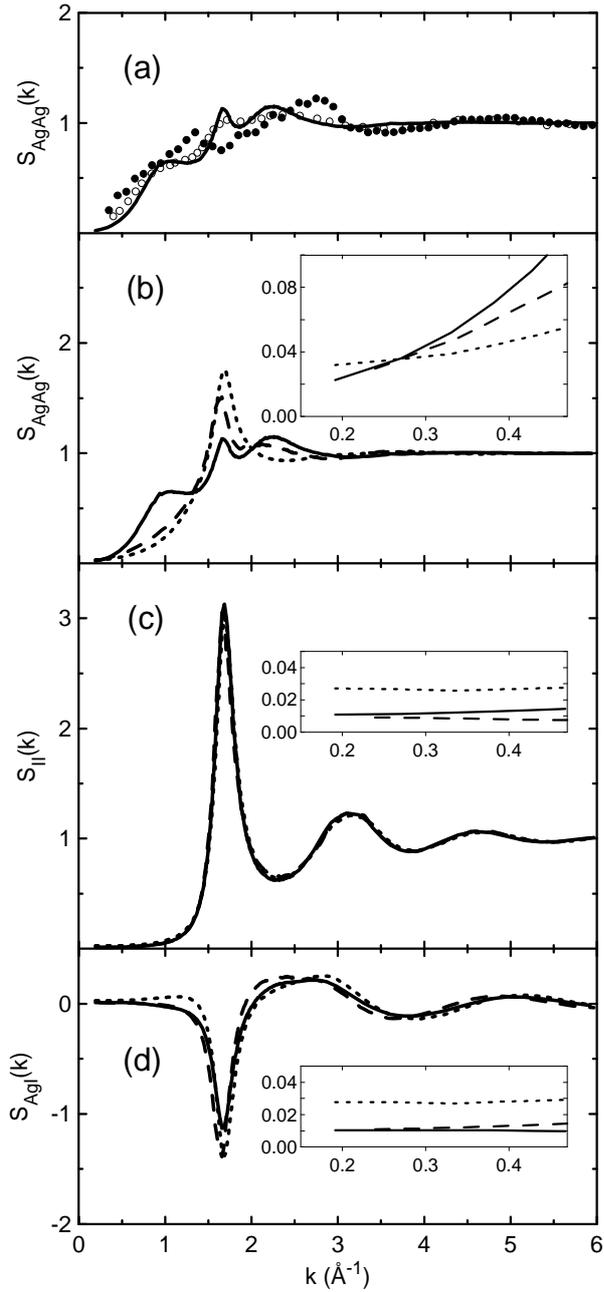

**Figure 5.** Ashcroft-Langreth partial structure factors $S_{AgAg}(k)$ (b), $S_{II}(k)$ (c), and $S_{AgI}(k)$ (d), from MD simulations of molten AgI at 923 K using the RIM (dotted line), PIM1 (solid line), and PIM2s (dashed line). The insets show the long-wavelength behavior. In panel (a) the $S_{AgAg}(k)$ for PIM1 is compared with those by Shimojo et al.[25] (open circles) and Kawakita et al.[49] (solid circles).

The long-wavelength behavior of the partial structure factors is related to the degree of compactness of the ionic packing, and their limit values $S_{ab}(0)$ are proportional to the isothermal compressibility $c_T$.[40,50] For monohalides the limit value is the same for the three partials, $S_{ab}(0) = 0.5 r_N k_B T c_T$, where $k_B$ is the Boltzmann's constant and $T$ the temperature. As can be seen in the insets of Figure 5, at the lowest $k$ accessible in our MD simulations the $S_{ab}(k)$ for PIM1 and PIM2s go to the same value, which is lower than that for RIM. The RIM compressibility is higher because the corresponding pair potential $f_{ab}^{VR}(r)$ includes the attractive contribution $-P_{ab}/r^4$ and, as it has been discussed above, the effective size of the RIM iodides is slightly smaller, and they are less closely packed, than in the polarizable models. This conjecture has been confirmed by a complementary MD simulation of a rigid ion model with $f_{ab}^{0}(r)$. This simulation shows that the corresponding $S_{ab}(k)$ go to the same limit value as those for PIM1 and PIM2s. It is worth noting that the values of $S_{II}$ and $S_{AgI}$ at the lowest wave numbers are close to the long-wavelength limit, while the values of $S_{AgAg}$ are higher and decay as $k$ approaches zero. We interpret this result as follows. Because the silver ions are smaller than iodides, they are less packed. This effect is reinforced for PIM1 and PIM2s since the induced polarization favors a less marked silver ion structure.

The prepeak in $S_{AgAg}(k)$ leads to a prepeak in the coherent static structure factor $S(k)$, which has also been observed experimentally by the Kyushu University group.[36,49] Neutron diffraction data by Shirakawa et al.[48] for molten copper halides also show a similar feature at ~1 Å$^{-1}$. In Figure 6 we compare the $S(k)$ for the three models and that from neutron diffraction data. The three MD $S(k)$ present a relatively structureless form as a result of subtle cancellations between the $S_{ab}(k)$. They present a broad main peak between 1.7 Å$^{-1}$ and 3.0 Å$^{-1}$ with the maximum at an intermediate $k$ between the second peak of $S_{II}$ and the first $S_{AgI}$ maximum beyond the first valley. The maxima for the three models are slightly displaced towards a higher $k$ (2.9 Å$^{-1}$) with respect to the scattering data, and the oscillations beyond the main peak damp faster than experiment. The PIM1 results are the closest to experiment, with the maximum value in agreement, although the mid shoulder at 2.25 Å$^{-1}$, related to the second peak of $S_{AgAg}$ discussed above, is not present in the neutron scattering data. While the $S(k)$ for RIM and PIM1 exhibit a small peak at $k_M \approx 1.7$ Å$^{-1}$, where experimental data show a shoulder, the first maximum of the $S(k)$ for PIM2s is at 2.1 Å$^{-1}$, the same $k$ of the PIM2s $S_{AgAg}$ second maximum.

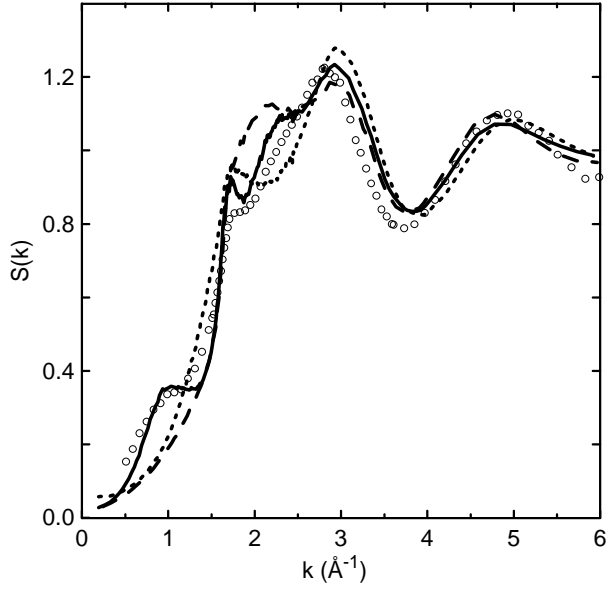

Figure 6. Coherent static structure factor $S(k)$ of molten AgI at 923 K from neutron scattering data (open circles),[36] and MD simulations using the RIM (dotted line), PIM1 (solid line), and PIM2s (dashed line).

Since the values of coherent scattering lengths $b_{Ag}$ and $b_I$ are not very different, $S(k)$ resembles $S_{NN}(k) = [S_{++}(k)+S_{--}(k)+2S_{+-}(k)]/2$ and, thus, provides information about the topological order of the melt, namely the correlations between ions irrespectively of their species. The chemical order is described by the charge static structure factor $S_{ZZ}(k) = [S_{++}(k)+S_{--}(k)-2S_{+-}(k)]/2$. The results for the three models are shown in Figure 7. The pronounced main peak at $k_M \approx 1.7$ Å$^{-1}$, less pronounced for PIM1, is a clear manifestation of the charge ordering. As it can be seen in the inset of Figure 7, the RIM $S_{ZZ}(k)$ at $k < 0.4$ Å$^{-1}$ practically fits the theoretical long-wavelength limit approximation derived for rigid ion models, i.e., $S_{ZZ}(k \to 0) = k^2/k_D^2$, where $k_D = (4\pi z^2 e^2 r_N/k_B T)^{1/2}$ is the Debye wave number with $z = z_+ = |z_-|$.[40,51] However, the corresponding values for PIM1 and PIM2s are higher than those for RIM, with those for PIM1 increasing much faster because its $S_{AgAg}(k)$ goes to its shoulder at about 1 Å$^{-1}$. Recently we have derived theoretically that the long-wavelength limit for ionic systems with point dipoles, as PIM1 and PIM2s, is given by $S_{ZZ}(k \to 0) = [1-C(k)]k^2/k_D^2$, where $C(k)$ is related to spatial correlations between charge and dipole moment densities.[52] As can be seen in the inset of Figure 7, the three models fulfill their theoretical $k \rightarrow 0$ limits derived in ref 52.

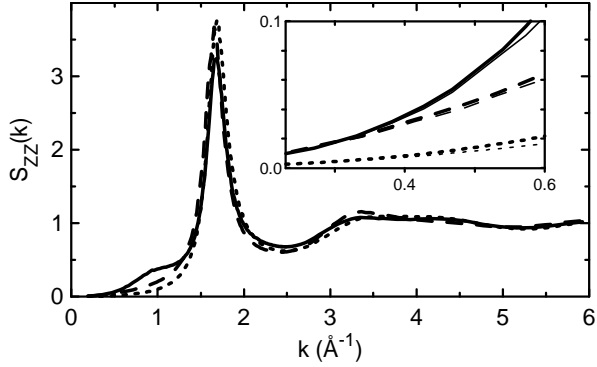

**Figure 7.** Charge static structure factors, $S_{ZZ}(k)$, from MD simulations of molten AgI at 923 K using the RIM (dotted line), PIM1 (solid line), and PIM2s (dashed line). The inset shows the long-wavelength behavior. The thin lines in the inset are the theoretical $k \rightarrow 0$ behavior for the three models.

**4.3. Ionic transport properties.** In Table 2 we present the MD results of the self-diffusion coefficients, $D_a$, and the ionic conductivities, $s$, for $a$-AgI at 573 K and molten AgI at 923 K. In the superionic $a$-phase the iodides remain around their lattice sites, and thus $D_I = 0$, whereas the silver ions show a liquid-like behavior and diffuse through the essentially rigid framework formed by the iodides. The three models retain in the liquid phase the superionic character of $a$-AgI, as the diffusivity of the silver ions is significantly larger than that for the iodides. Since silver ions are smaller than iodides, they can diffuse through the packed structure of the slowly diffusing iodides. The induced polarization effects are clearly observed on the silver ion diffusivity for both the liquid and $a$-phase. The value of $D_{Ag}$ for PIM1 is larger than that for RIM, with that for PIM2s in between. The induced polarization on the iodides allows the small silver ions to move more easily through the interstices of the iodides structure, that is, it increases the free space for the cations that can diffuse through "faster" channels. On the contrary, the value of $D_I$ for liquid PIM1 is lower than that for RIM, with that for PIM2s in between.

Although it could be a polarization effect, we believe that this result is mainly related to the small changes of the effective iodide size discussed above in going from the pair potential $f_{ab}^{VR}(r)$ used for RIM to the $f_{ab}^{0}(r)$ for the polarizable ion models. Anyway, the $D_I$ in the liquid is similar in the three models. Unfortunately, we are not aware of experimental data for the self-diffusion coefficients, except the $D_{Ag}$ in the $a$-phase given by Kvist and Tarneberg,[53] which is near the RIM value (see Table 2).

TABLE 2: MD results of the self-diffusion coefficients, $D_a$, and the ionic conductivity, $s$, for $a$-AgI at 573 K and molten AgI at 923 K.

|  | $a$-AgI (573 K) | | | Molten AgI (923 K) | | |
| --- | --- | --- | --- | --- | --- | --- |
|  | RIM | PIM2s | PIM1 | RIM | PIM2s | PIM1 |
| $D_{Ag}$ / $10^{-5}$ cm$^2$/s | 2.34 | 2.70 | 3.02 | 6.50 | 7.24 | 9.08 |
| $D_I$ / $10^{-5}$ cm$^2$/s | 0 | 0 | 0 | 2.22 | 1.88 | 1.75 |
| $s$ / $(\Omega\text{cm})^{-1}$ | 2.2 | 2.1 | 1.9 | 3.4 | 2.8 | 2.8 |
| $s_{NE}$ / $(\Omega\text{cm})^{-1}$ | 1.2 | 1.3 | 1.5 | 2.5 | 2.6 | 3.1 |
| $\Delta$ | −0.87 | −0.53 | −0.27 | −0.36 | −0.08 | 0.10 |

Experimental $s$ values: $s$(573 K) = 1.97 $(\Omega\cdot\text{cm})^{-1}$ (ref 8), and $s$(923 K) = 2.51 $(\Omega\cdot\text{cm})^{-1}$ (ref 37). Experimental $D_{Ag}$ value: $D_{Ag}$(573 K) = 2.41·$10^{-5}$ cm$^2$/s (ref 53).

The normalized self-velocity autocorrelation functions, $C_a(t)$, plotted in Figure 8 for the molten phase help to understand the ionic diffusion discussed above. The $C_{Ag}(t)$ for the three models shows a typical diffusive behavior, with a slow decay and a weak minimum, in contrast with the oscillatory behavior of the $C_I(t)$. As it was suggested in ref 54 for copper and silver halide melts, while the short-time microscopic motion of the small cations is solely determined by the first neighboring shell of unlike ions, the large anions are so closely packed that they experience a rattling motion within a double cage formed by unlike and like first neighbors.

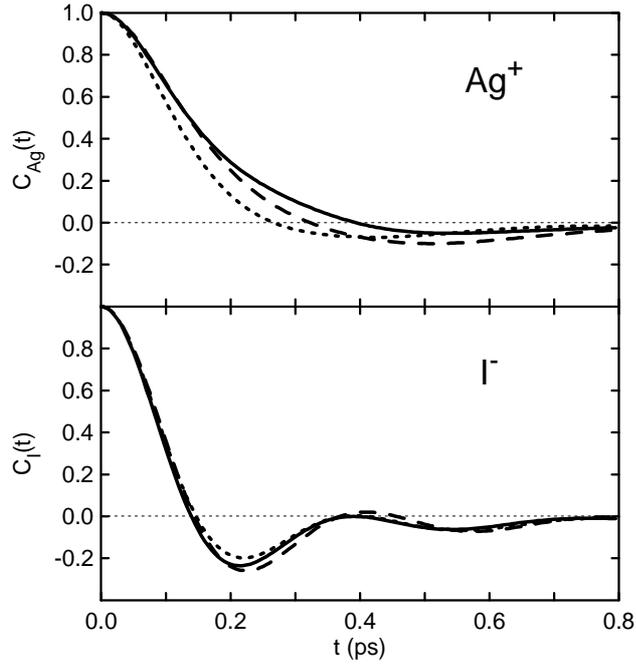

**Figure 8.** Normalized self-velocity autocorrelation functions, $C_{Ag}(t)$ (top) and $C_I(t)$ (bottom), from MD simulations of molten AgI at 923 K, using the RIM (dotted line), PIM1 (solid line), and PIM2s (dashed line).

Since the induced polarization on the anions increases the free space for the cations, the $C_{Ag}(t)$ initial decay is considerably slower for the two polarizable ion models than for the RIM, and the weak minimum is at longer $t$. The short-range damping polarization effects cancel partially this effect for PIM2s, whose $C_{Ag}(t)$ has a minimum deeper than that for RIM. The $C_I(t)$ for the three models exhibit a pronounced backscattering minimum at about the same value of $t$, and subsequent damped oscillations with approximately the same frequency. The variation in the depth of the minimum is related to the differences between the pair potentials $f_{ab}^{VR}(r)$ and $f_{ab}^{0}(r)$. Since the absence of the $P_{ab}$ terms in $f_{ab}^{0}(r)$ increases the short-range repulsion between anions, the backscattering is more pronounced for the polarizable models, consistently with the conclusions drawn in ref 54. Moreover, the reduction of the PIM1 minimum depth with respect to that for PIM2s is attributable to the non-damped polarization effects in the former model. The $C_a(t)$ obtained from the simulations of the *a*-phase are qualitatively similar to those of the liquid, although the minimum of $C_{Ag}(t)$ is deeper, and at a slightly shorter $t$, in the solid case, and the oscillations of $C_I(t)$ are considerably more pronounced.

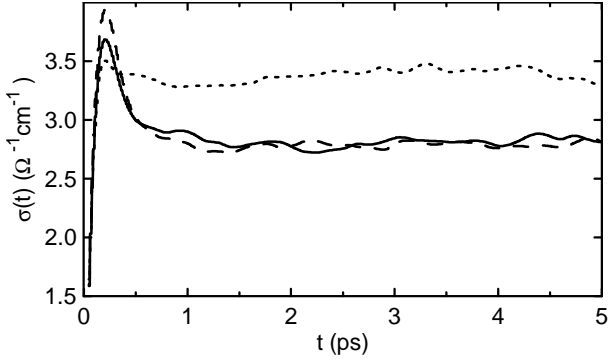

**Figure 9.** Values of the Kubo integral for the conductivity as function of the upper limit time, $s(t) = (r_N e^2 / k_B T) \int_0^t \Lambda_{ZZ}(t')dt'$, from MD simulations of molten AgI at 923 K, using the RIM (dotted line), PIM1 (solid line), and PIM2s (dashed line).

From the charge-density current autocorrelation function,

$$\Lambda_{ZZ}(t) = \frac{1}{3} N \langle \mathbf{v}_Z(t) \cdot \mathbf{v}_Z(0) \rangle \quad \text{with} \quad \mathbf{v}_Z = \frac{1}{N} \sum_{i=1}^N z_i \mathbf{v}_i , \qquad (19)$$

and the mean square displacement of the charge-center position,

$$Q_{ZZ}(t) = \frac{1}{3} N \langle |\mathbf{r}_Z(t) - \mathbf{r}_Z(0)|^2 \rangle \quad \text{with} \quad \mathbf{r}_Z = \frac{1}{N} \sum_{i=1}^N z_i \mathbf{r}_i , \qquad (20)$$

where we assume $|z_i| = 1$, we have evaluated the ionic conductivity $s$ by the corresponding Kubo and Einstein-like relations,

$$s = \frac{r_N e^2}{k_B T} \int_0^\infty \Lambda_{ZZ}(t)dt \quad \text{and} \quad s = \frac{r_N e^2}{k_B T} \lim_{t \to \infty} \frac{Q_{ZZ}(t)}{2t} . \qquad (21)$$

The estimated uncertainty for the $s$ values reported on Table 2 is less than 5%, greater than those for $D_a$ (less than 1%) because $s$ is a collective property that requires to be averaged over long simulations. In Figure 9 it can be seen the time evolution of $s(t) = (r_N e^2 / k_B T) \int_0^t \Lambda_{ZZ}(t')dt'$ obtained for the three models at the molten phase, and its oscillating convergence at times longer than 2 ps. The uncertainty has been estimated from the fluctuating values of $s(t)$.

Some features are common for both the $a$- and molten phases results. The conductivity for polarizable ion models is smaller than that for RIM, and in better agreement with experiment (see Table 2). In Figure 10 it can be seen that the $Q_{ZZ}(t)$ slopes for PIM1 and PIM2s at the molten phase are very similar, and clearly lower than that for RIM, in accordance with the converging values of $s(t)$.

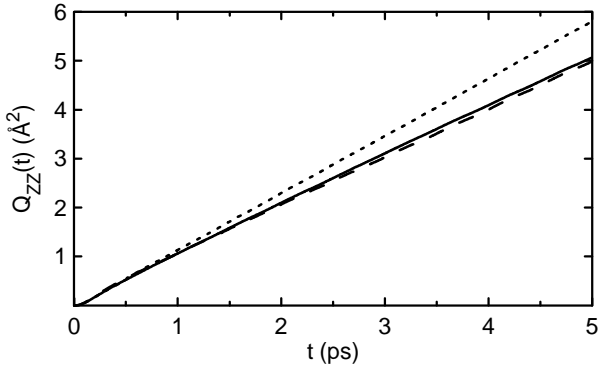

**Figure 10.** Mean square displacement of the charge-center position, $Q_{ZZ}(t)$, from MD simulations of molten AgI at 923 K, using the RIM (dotted line), PIM1 (solid line), and PIM2s (dashed line).

The larger silver ion diffusivity for PIM1 does not lead to a larger $s$ as would be expected from the Nernst-Einstein approximation, which for monohalides can be written as $s_{NE} = r_N e^2 (D_{Ag} + D_I)/(2k_B T)$.[51] On the contrary, the PIM1 conductivity has the lowest value in both phases in spite of presenting the highest silver ion diffusivity. $s$ is not equal to $s_{NE}$ because, besides the self-diffusion term $s_{NE}$, there is a contribution $-\Delta s_{NE}$ due to the correlations between distinct ions, i.e., $s = s_{NE}(1-\Delta)$. In all simulated cases, except molten PIM1, $s$ is higher than $s_{NE}$ and, thus, $\Delta < 0$. Moreover the values of $s$ for the three models in each phase are ordered from lower to higher in the inverse order of $s_{NE}$ (and $\Delta$), showing the different role played by distinct correlations. From the experimental values of $s$, 1.97 $(\Omega \cdot cm)^{-1}$,[8] and $D_{Ag}$, $2.41\cdot 10^{-5}$ cm$^2$/s,[53] for *a*–AgI at 573 K, it is derived that $\Delta = -0.6$, which is closer to the PIM2s result. We are not aware of experimental data of $D_{Ag}$, and thus of $\Delta$, for molten AgI. The only experimental result on $\Delta$ that we know for a superionic melt is $\Delta = 0.2$ for molten CuCl,[55] which is positive as for the PIM1 result. We recall that in molten alkali halides the experimental values of $\Delta$ are also positive.[56]

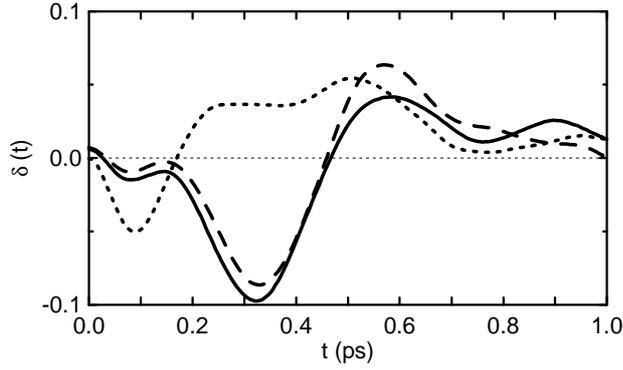

**Figure 11.** Normalized distinct velocity correlation function $d(t)$ from MD simulations of molten AgI at 923 K using the RIM (dotted line), PIM1 (solid line), and PIM2s (dashed line).

The distinct contribution to $\Lambda_{ZZ}(t)$ for AgI can be written as

$$\Lambda^d_{ZZ}(t) = \Lambda_{ZZ}(t) - \tfrac{1}{2}[\Lambda_{Ag}(t) + \Lambda_I(t)], \quad (22)$$

where $\Lambda_{Ag}(t)$ and $\Lambda_I(t)$ are the (not normalized) self-velocity autocorrelation functions, the parameter $\Delta$ is proportional to the area under $\Lambda^d_{ZZ}(t)$,

$$\Delta = -\frac{2}{D_{Ag} + D_I}\int_0^\infty \Lambda^d_{ZZ}(t)dt. \quad (23)$$

In Figure 11 we present $d(t) = \Lambda^d_{ZZ}(t)/\Lambda_{ZZ}(0)$ obtained from the MD simulations of the molten phase. While the $d(t)$ for RIM is negative between 0 and 0.2 ps and positive between 0.2 and 0.7 ps, those for the two polarizable ion models show the same qualitative behavior with a negative minimum at about 0.33 ps and a positive maximum at around 0.58 ps. Nevertheless, $\Delta$ is positive as the area under $\mathbf{d}(t)$ is negative, and $\mathbf{s} < \mathbf{s}_{NE}$, for PIM1, whereas $\Delta$ is negative and $\mathbf{s} > \mathbf{s}_{NE}$ for PIM2s.

## 5. Concluding remarks

In this work we have carried out molecular dynamics (MD) simulations for three models of AgI, a rigid ion model (RIM) with the effective interionic pair potential of the Vashishta and Rahman form and the parameterization proposed by Shimojo,[19] and two polarizable ion models (PIM1 and PIM2s). The latter are based on the RIM pair potential, to which the polarization effects are added. In PIM1 only the iodides are assumed to be polarizable and the dipole moment of each iodide is induced by the local electric field due to all other ions. In PIM2s both species are polarizable, and the dipoles are not only induced by the field but also by the short-range overlap effects due to the neighbors. In order to guarantee the stability of the simulations, the short-range damping polarization must oppose the dipole contributions of the field created by both the charges and dipoles of other ions. Under these conditions, we solve the "polarization catastrophe" problem, which had prevented the simulation of a similar model in our previous work on AgBr.[23]

The solid $\alpha$-phase of AgI has been simulated at 573 K. The three models reproduce the superionic character of $\alpha$-AgI. The averaged spatial distribution of the diffusive silver ions within the bcc iodide sublattice is qualitatively similar in the three models, and reproduces the well-known experimental results. The cations tend to be located around the tetrahedral sites and, to a lesser extent, around trigonal and octahedral sites. However, while in the RIM and PIM2s the separation between neighboring cations is maximized (4.3 Å), the PIM1 $g_{AgAg}$ peaks at 3.2 Å, which means that in this model two nearest silver ions tend to locate around tetrahedral sites linked by $(1/2,1/4,1/4)a$ or equivalent vectors of the same modulus. The reverse Monte Carlo analysis by Nield et al.[5] and the ab-initio MD calculations by Wood and Marzari[45] agree on the existence of a peak at around 3 Å.

Regarding the liquid, we have found that the iodide structure does not depend significantly on the polarization effects. On the contrary, it is mainly the silver ions that are affected, which can be attributed to the screening of the cations repulsion due to the induced dipoles on the iodides. The silver ion structure for the two polarizable models is less marked, and the cations penetration within the first shell deeper, than for the RIM. Furthermore, as in the solid state simulations, the distance between nearest cations is shorter for PIM1, and to a lesser extent for PIM2s, with respect to RIM. As a result of the induced polarization, the silver ions have more free space to move, and tend to diffuse through the interstices of the iodides structure, in such a way that the silver ions self-diffusion coefficients in both the *a*- and molten phases are higher for PIM1 than for PIM2s and RIM (in this order). In spite of this fact, and due to the distinct velocity correlations, the ionic conductivity is lower for the polarizable models than that for RIM. The superionic character of the melt in the three models is reflected in the significantly larger mobility of the cations compared to that of anions.

The static coherent structure factor $S(k)$ has been calculated for the liquid and compared to experimental data. The three models present a broad main peak between 1.7 Å$^{-1}$ and 3.0 Å$^{-1}$, similar to that from neutron scattering experiments, but the PIM1 $S(k)$ is the only one that reproduces the experimental prepeak at about 1 Å$^{-1}$. The prepeak is exhibited also by the PIM1 Ag-Ag partial structure factor. We have suggested that it is due to the inhomogeneous distribution in space of silver ions, as a consequence of polarization effects.[24] The closer approach between neighboring silver ions observed in PIM1 causes the appearance of low Ag$^+$ density zones, whose periodicity gives rise to the prepeak. Moreover, the PIM1 and PIM2s conductivity is in better agreement with experimental values, and the PIM1 deviation parameter Δ is positive for molten AgI, like the experimental results for molten alkali halides and molten CuCl.

The overall results for PIM1 present a better agreement with experimental data than the other two models. In particular, the appearance of the prepeak in $S(k)$ leads us to conclude that the anions induced polarization plays an important role in the real system. However, some improvements must be made in the models in order to advance in the understanding of this system. The next steps on modelling AgI may be the correction of the pair potential, and the simulation of a model which takes into account the polarizability of both species, but canceling the polarization effects to a lesser extent than in PIM2s.

**Acknowledgements.** This work was supported by DGI of Spain (Grant No. FIS2006-12436-C02-01), the DURSI of the Generalitat of Catalonia (Grant No. 2005SGR-00779) and European Union FEDER funds (Grant No. UNPC-E015). One of us (VB) also thanks the Spanish Ministry of Education and Science for FPU Grant No. AP2003-3408.

**References**

(1) Keen, D. A. *J. Phys.: Condens. Matter* **2002**, *14*, R819.

(2) Burley, G. *J. Phys. Chem.* **1964**, *68*, 1111.

(3) Hoshino, S.; Sakuma, T.; Fujii, Y. *Solid State Commun.* **1977**, *22*, 763.

(4) Cava, R. J.; Reidinger, F.; Wuensch, B. J. *Solid State Commun.* **1977**, *24*, 411.

(5) Nield, V. M.; Keen, D. A.; Hayes, W.; McGreevy, R. L. *Solid State Ionics* **1993**, *66*, 247.

(6) Cava, R. J.; Fleming, R. M.; Rietman, E. A. *Solid State Ionics* **1983**, *9-10*, 1347.

(7) Boyce, J. B.; Hayes, T. M.; Stutius, W.; Mikkelsen Jr., J. C. *Phys. Rev. Lett* **1977**, *38*, 1362.

(8) Tubandt, C.; Lorenz, E. *Z. Phys. Chem.* **1914**, *87*, 513.

(9) Vashishta, P.; Rahman, A. *Phys. Rev. Lett.* **1978**, *40*, 1337.

(10) Parrinello, M.; Rahman, A.; Vashishta, P. *Phys. Rev. Lett.* **1983**, *50*, 1073.

(11) Tallon, J. L. *Phys. Rev. B* **1988**, *38*, 9069.

(12) Chiarotti, G. L.; Jacucci, G.; Rahman, A. *Phys. Rev. Lett.* **1986**, *57*, 2395.

(13) Madden, P. A.; O'Sullivan, K. F.; Chiarotti, G. *Phys. Rev. B* **1992**, *45*, 10206.

(14) Howe, M. A.; McGreevy, R. L.; Mitchell, E. W. J. *Z. Phys. B* **1985**, *62*, 15.

(15) Stafford, A. J.; Silbert, M. *Z. Phys. B* **1987**, *67*, 31.

(16) Takahashi, H.; Takeda, S.; Harada, S.; Tamaki, S. *J. Phys. Soc. Jpn.* **1988**, *57*, 562.

(17) Stafford, A. J.; Silbert, M.; Trullàs, J.; Giró, A. *J. Phys.: Condens. Matter* **1990**, *2*, 6631.

(18) Trullàs, J.; Giró, A.; Silbert, M. *J. Phys.: Condens. Matter* **1990**, *2*, 6643.

(19) Shimojo, F.; Kobayashi, M. *J. Phys. Soc. Jpn.* **1991**, *60*, 3725.

(20) Madden, P. A.; Wilson, M. *Chem. Soc. Rev.* **1996**, *25*, 339.

(21) Wilson, M.; Madden, P. A.; Costa-Cabral, B. J. *J. Phys. Chem.* **1996**, *100*, 1227.

(22) Trullàs, J.; Alcaraz, O.; González, L. E.; Silbert, M. *J. Phys. Chem. B* **2003**, *107*, 282.

(23) Bitrián, V.; Trullàs, J. *J. Phys. Chem. B* **2006**, *110*, 7490.

(24) Bitrián, V.; Trullàs, J.; Silbert, M. *J. Chem. Phys.* **2007**, *126*, 021105.

(25) Shimojo, F.; Inoue, T.; Aniya, M.; Sugahara, T.; Miyata, Y. *J. Phys. Soc. Jpn.* **2006** *75*, 114602.

(26) Wilson, M.; Madden, P. A. *J. Phys.: Condens. Matter* **1993**, *5*, 2687.

(27) Rick, S. W.; Stuart, S. J. In *Reviews in Computational Chemistry*; Lipkowitz, K. B., Boyd, D. B., Eds.; John Wiley and Sons: New York, 2002**;** Volume 18.

(28) Hutchinson, F.; Wilson, M.; Madden, P. A. *Mol. Phys.* **2001**, *99*, 811; Wilson, M.; Madden, P. A. *J. Phys.: Condens. Matter* **1994**, *6*, 159.

(29) Thole, B. T. *Chem. Phys.* **1981**, *59*, 341.

(30) Burnham, C. J.; Li, J.; Xantheas, S. S.; Leslie, M. *J. Chem. Phys.* **1999**, *110*, 4566.

(31) Tang, K. T.; Toennies, J. P. *J. Chem. Phys.* **1984**, *80*, 3726.

(32) Ahlström, P.; Wallqvist, A.; Engström, S.; Jönsson, B. *Mol. Phys.* **1989**, *68*, 563.

(33) Böttcher, C. J. F. *Theory of Electric Polarization,* 2nd ed.; Elsevier: Amsterdam, 1993.

(34) Bucher, M. *Phys. Rev. B* **1984**, *30*, 947.

(35) Alcaraz, O.; Bitrián, V.; Trullàs, J. *J. Chem. Phys.* In press.

(36) Bitrián, V.; Trullàs, J.; Silbert, M.; Enosaki, T.; Kawakita, Y.; Takeda, S. *J. Chem. Phys.* **2006**, *125*, 184510.

(37) Janz, G. J.; Dampier, F. W.; Lakdhminaryanah, G. R.; Lorentz, P. K.; Tomkins, R. P. T. *Molten Salts*, National Bureau of Standards Reference Data Series, Washington DC, 1968; Vol. 15.

(38) Vesely, F. J. *J. Comput. Phys.* **1977**, *24*, 361.

(39) Ashcroft, N. W.; Langreth, D. C. *Phys. Rev.* **1967**, *156*, 685.

(40) Rovere, M.; Tosi, M. P. *Rep. Prog. Phys.* **1986**, *49*, 1001.

(41) Sears, V. F. *Neutron News* **1992**, *3*, 26.

(42) Tosi, M. P.; Price, D. L.; Saboungi, M. -L. *Annu. Rev. Phys. Chem.* **1993**, *44*, 173.

(43) Gray-Weale, A.; Madden, P. A. *Mol. Phys.* **2003**, *101*, 1761.

(44) Chahid, A.; McGreevy, R. L. *J. Phys.: Condens. Matter* **1998**, *10*, 2597.

(45) Wood, B. C.; Marzari, N. *Phys. Rev. Lett.* **2006**, *97*, 166401.

(46) Eisenberg, S.; Jal, J. F.; Dupuy, J.; Chieux, P.; Knoll, W. *Phil. Mag. A* **1982**, *46*, 195.

(47) Nield, V. M.; McGreevy, R. L.; Keen, D. A.; Hayes, W. *Physica B* **1994**, *202*, 159.

(48) Shirakawa, Y.; Saito, M.; Tamaki, S.; Inui, M.; Takeda, S. *J. Phys. Soc. Jpn.* **1991**, *60*, 2678.

(49) Kawakita, Y.; Tahara, S.; Fujii, H.; Kohara, S.; Takeda, S. *J. Phys.: Condens. Matter* **2007**, *19*, 335201.

(50) McGreevy, R. L. *Sol. St. Phys.* **1987**, *40*, 247.

(51) Hansen, J. P.; McDonald, I. R. *Theory of Simple Liquids*, 2nd ed.; Academic: London, 1986.

(52) Bitrián, V.; Trullàs, J. *J. Phys: Conf. Series*. Submitted for publication.

(53) Kvist, A.; Tarneberg, R. *Z. Naturforsch.* **1970**, *25A*, 257.

(54) Alcaraz, O.; Trullàs, J. *J. Chem. Phys.* **2000**, *113*, 10635.

(55) Poignet, J. C.; Barbier, M. J. *Electrochim. Acta* **1981**, *26*, 1429.

(56) Ciccotti, G.; Jacucci, G.; McDonald, I. R. *Phys. Rev. A* **1976**, *13*, 426.